# ACCELERATING SCIENCE: A COMPUTING RESEARCH AGENDA

Vasant G. Honavar[1], Mark D. Hill[2], and Katherine Yelick[3]
Computing Community Consortium
2/19/2016

Version 1

**Abstract**

The emergence of "big data" offers unprecedented opportunities for not only accelerating scientific advances but also enabling new modes of discovery. Scientific progress in many disciplines is increasingly enabled by our ability to examine natural phenomena through the computational lens, i.e., using algorithmic or information processing abstractions of the underlying processes; and our ability to acquire, share, integrate and analyze disparate types of data. However, there is a huge gap between our ability to acquire, store, and process data and our ability to make effective use of the data to advance discovery. Despite successful automation of routine aspects of data management and analytics, most elements of the scientific process currently require considerable human expertise and effort. Accelerating science to keep pace with the rate of data acquisition and data processing calls for the development of algorithmic or information processing abstractions, coupled with formal methods and tools for modeling and simulation of natural processes as well as major innovations in *cognitive tools* for scientists, i.e., computational tools that leverage and extend the reach of human intellect, and partner with humans on a broad range of tasks in scientific discovery (e.g., identifying, prioritizing formulating questions, designing, prioritizing and executing experiments designed to answer a chosen question, drawing inferences and evaluating the results, and formulating new questions, in a closed-loop fashion). This calls for concerted research agenda aimed at: Development, analysis, integration, sharing, and simulation of algorithmic or information processing abstractions of natural processes, coupled with formal methods and tools for their analyses and simulation; Innovations in cognitive tools that augment and extend human intellect and partner with humans in all aspects of science. This in turn requires: the formalization, development, analysis, of algorithmic or information processing abstractions of various aspects of the scientific process; the development of computational artifacts (representations, processes, protocols, workflows, software) that embody such understanding; and the integration of the resulting cognitive tools into collaborative human-machine systems and infrastructure to advance science.

**OVERVIEW**

Tycho Brahe gathered considerable and accurate data on the movement of the planets ("big data" for his time). However, this data did not find real value until Johannes Kepler used it to discover his three laws of planetary motion. Later Isaac Newton used these laws and other data to derive his unified laws of motion, and lay the foundations of classical physics. To do so, he had to invent calculus for describing such things as rates of change. Brahe, Kepler, and Newton were all engaged in the practice of science, a systematic process for acquiring knowledge through observation or experimentation and developing theories to describe and explain natural phenomena. The past centuries have witnessed major scientific breakthroughs as a result of advances in instruments of observation, formalisms for describing the laws of nature, and improved tools for calculation.

---

[1] Professor and Edward Frymoyer Chair, College of Information Sciences and Technology, Professor of Computer Science, Bioinformatics and Computational Biology and Neuroscience Graduate Programs, Director, Center for Big Data Analytics and Discovery Informatics, and Associate Director, Institute for Cyberscience, Pennsylvania State University, University Park, PA 16802. Email: vhonavar@ist.psu.edu; Web: http://faculty.ist.psu.edu/vhonavar

[2] John P. Morgridge and Gene P. Amdahl Professor and Chair, Department of Computer Sciences, University of Wisconsin-Madison, Madison, WI 53706. Email: markhill@cs.wisc.edu; Web: http://pages.cs.wisc.edu/~markhill/

[3] Professor, Computer Science Division, University of California at Berkeley, Berkeley, CA 94720 and Associate Laboratory Director of Computing Sciences, Lawrence Berkeley National Laboratory. Email: yelick@berkeley.edu; Web: http://www.cs.berkeley.edu/~yelick/

Today, the experimental instruments are more powerful, the scientific questions more complex, and the mathematical, statistical and computational methods for analyzing data have become more sophisticated. The resulting emergence of "big data" offers unprecedented opportunities for accelerating science. Arguably, "big data" accelerates Brahe's part of the scientific endeavor, and increasingly, Kepler's part, with the increasing use of machine learning for building models from data. Nevertheless, many aspects of the scientific process (designing, prioritizing and executing experiments, organizing data, integrating data, identifying patterns, drawing inferences and interpreting results) constitute an even greater bottleneck than ever.

The goal of this white paper is to articulate a research agenda for developing cognitive tools that can augment human intellect, and partner with humans on all aspects of the scientific process, including in particular, those that are exacerbated by "big data." We argue that there is great opportunity for dramatically accelerating science and enabling new modes of scientific discovery, perhaps even empowering and enabling the future likes of Kepler or Newton in the era of big data.

The benefits of accelerating science extend well beyond the scientific community to all of humanity. Imagine: Precision health regimens that take into account not only one's genetic makeup, but also environment, and lifestyle; Personalized education that optimizes curriculum, pedagogy, etc. to optimize the learning outcomes for each individual; Precision agriculture that optimizes everything from the choice of crops to water and fertilizer use to optimize yield and impact on the environment. These are just a start, however, as in the 21st century we should be able to invent technologies undreamed of in this century's early years, as who in 1900 could have anticipated 20th century advances, such as the Internet (no computers yet) or DNA sequencing (DNA structure unknown)?

**ACCELERATING SCIENCE: THE VALUE PROPOSITION**

Imagine a world in which scientists work with cognitive tools that can

- Given access to literature and data:
  - Create and share a knowledge base that summarizes what we know about a scientific question (annotated with uncertainty, provenance, and underlying assumptions);
  - Summarize and prioritize questions that need to be answered to achieve an overall scientific objective (e.g., understanding the molecular mechanisms that underlie cancer);
  - Identify and rank alternative explanations of an observation based on the current state of scientific understanding in a given field;
  - Design and prioritize study techniques;
  - Construct a computational model, e.g., a network of genes that orchestrate a specific biological process of interest, that make experimentally testable predictions.
- Given a conjecture:
  - Identify data that support or refute the conjecture;
  - Identify simulations that can interpret the theory, e.g., over time or in various physical settings;
  - Design and prioritize, orchestrate, and execute experiments.

- Given an experimental design, experimental results, and access to literature:
  - Create a plan for replicating the study and validating the claims;
  - Generate and rank alternative interpretations of the data;
  - Document the study, communicate results;
  - Integrate results into the larger body of knowledge within or across disciplines.
- Given a collection of experimental and observational studies:
  - Infer a causal effect of interest, e.g., the role of a specific gene or combination of genes in a specific biological process;
  - Calculate scientific parameters, e.g., geophysical characteristics affecting earthquakes, by solving an inverse problem by comparing simulations to the observations.
- Given a scientific question and a network of researchers, assemble a team that is best equipped to answer the question.
- Track scientific progress, evolution of scientific disciplines, and scientific impact.

Cognitive tools for acclerating science could lead to dramatic increases in scientific productivity by increasing efficiency of the key steps in scientific process, and in the quality of science that is carried out (by reducing error, enhancing reproducibility), allow scientific treatment of topics that were previously impossible to address, and enable new modes of discovery that leverage large amounts of data, knowledge, and automated inference. The sections that follow attempt to further flesh out our vision for accelerating science by accelerating increasingly larger fractions of the scientific process.

## THE SCIENTIFIC PROCESS

To understand where the major bottlenecks to scientific progress are, it is useful to revisit a simplified model or template of the scientific process (See Hacking, 1983; Chalmers, 1999; Rosenberg, 2000 for reviews). Figure 1 summarizes the key elements of the scientific process.

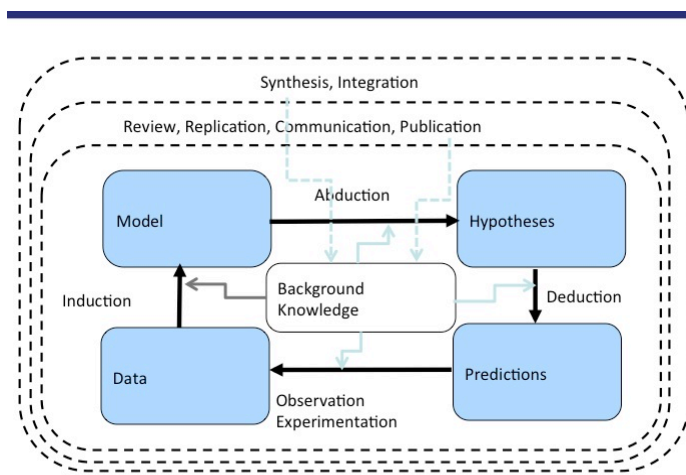

Figure 1: A Cartoon of the Scientific Process

Typically, scientific inquiry starts with a question within a domain of study, e.g., biology. With the question in hand, one has to assemble the background information and acquire the data necessary to answer the question. Then one proceeds to construct one or more models from data (and background information). Choosing a small set of models from among a much larger set of candidates involves additional considerations (simplicity, consistency with what else is known), etc. The models can be used to advance hypotheses that result, ideally, in testable predictions. The observations or experiments designed to test the predictions yield additional data that feed into the larger scientific process. Science is a social endeavor, with multiple individuals and teams, driven by intrinsic as well as extrinsic incentives. Scientific findings go through peer review, communication, and publication, and replication before they are integrated into the larger body of knowledge in the relevant discipline.

It is worth noting that there is considerable variability across scientific disciplines, e.g., in cosmology, where there is little possibility of executing designed experiments, one typically has to make do with observational data or the results of 'natural' experiments. Nevertheless, it is clear that the processes of acquiring, organizing, verifying, validating, integrating, analyzing, reasoning with, and communicating information (models, hypotheses, theories, explanations) about natural and built systems lie at the heart of the scientific enterprise.

**ACCELERATING SCIENCE: TRADITIONAL ENABLERS**

Major scientific advances are often enabled by:

- Advances in the instruments of observation (new measurement devices or methods or making it possible to acquire data of new modalities, higher resolution in time or space, or in larger volumes than previously possible).

- Development of mathematical models and methods for representing and reasoning about scientific hypotheses and theories (e.g., the invention of calculus by Newton and Leibnitz that were necessary for the advances in physics);

- Development of effective tools for data analysis and simulation (e.g., the invention of the computer that enabled among other things, solution of systems of linear equations, simulation of complex models of physical, biological, and cognitive processes);

- Cross-fertilization and integration of concepts, experimental methods, data, tools, hypotheses, theories, across disciplines (e.g., the emergence of molecular biology through convergence of biological and physical sciences).

In what follows, we argue that the emergence of big data and the ability to examine natural processes using the computational lens (Karp, 2011), offer the possibility of rapid acceleration of science. However, realizing this requires algorithmic or information processing abstractions of natural processes, coupled with formal methods and tools for their analyses and simulation; cognitive tools that augment and extend human intellect and partner with humans in all aspects of science.

**ACCELERATING SCIENCE: THE DRIVERS**

New technology in sensors, detectors, sequencing, imaging and simulation offers unprecedented opportunities for not only accelerating scientific advances, but also enabling new modes of scientific discovery. New scientific advances in many disciplines are increasingly being driven by our ability to acquire, share, integrate and analyze disparate types of data, leading to what has been suggested to be a new scientific paradigm, namely data-intensive science (Hey, Tansley, and Tolle, 2009). The resulting challenges in storage, organization, curation, access, sharing, management, processing, analytics, statistics, and visualization s are widely recognized and form the focus of much current research. Modern data analytics techniques that integrate sophisticated probabilistic models, statistical inference, and scalable data structures and algorithms into machine learning systems have resulted in powerful ways to extract actionable knowledge from data in virtually every area of human endeavor. Creative applications of data analytics are enabling biologists to gain insights into how living systems acquire, encode, process, and transmit information; neuroscientists to uncover the neural bases of cognition; health scientists to not only diagnose and treat diseases but also help individuals make healthy choices; economists to understand markets; physical scientists to improve our basic understanding of the physical world, security analysts to uncover threats to national security; social scientists to study the evolution and dynamics of social networks; and scholars to gain new understandings of literature, arts, history, and cultures through advances in the digital humanities. However, despite, and perhaps because of, advances in "big data" technologies for data acquisition, management and analytics, (bottom left of Figure 1), the other largely manual, and labor-intensive aspects the scientific process (the rest of Figure 1) have become the rate limiting steps in scientific progress.

Consider for example, the task of identifying a question for investigation in a domain of inquiry, e.g., the Life Sciences. This is a non-trivial task that requires a good grasp of the current state of knowledge, the expertise and skills needed, the instruments of observation available, the experimental manipulations that are possible, the data analysis and interpretation tools available, etc. Understanding the current state of knowledge requires mastery of the relevant scientific literature which, much like many other kinds of "big data", is growing at an exponential rate. For example, in 2011, the number of peer-reviewed biomedical research articles appearing in *Pubmed* exceeded 2700 articles per day. The sheer volume and the rate of growth of scientific literature makes it impossible for a scientist to keep up with advances that might have a bearing on the questions being pursued in his or her laboratory. The magnitude of this challenge is further compounded by the fact that many scientific investigations increasingly need to draw on data from a multitude of databases (e.g., Genbank, Protein Data Bank, etc. in the life sciences) and expertise and results from multiple disciplines.

As another example, consider the task of designing an optimal experiment that provides the most valuable information at the lowest cost to help answer a chosen scientific question requires a careful exploration of the space of possible experiments, their relative cost, risk, and feasibility, in the context of all that is known. This challenge is further compounded by the varying degrees of uncertainty associated with the scientific findings.

The components of the scientific process present similar challenges. This underscores the need for much improved cognitive tools tools for assisting scientists with the rate-limiting steps of the scientific process.

**ACCELERATING SCIENCE: FEASIBILITY**

In what follows, we argue that computation increasingly serves as:

➢ A language for science, a role not unlike that played by mathematics over the past many centuries; and

➢ A powerful formal framework and exploratory apparatus for the conduct of science. These developments together set the stage for developing the cognitive tools needed to accelerate science.

**Computing as a language of science**

It was nearly a century ago that Rutherford said "All science is either stamp collecting or physics". Advances in computing, storage, and communication technologies have made it possible to organize, annotate, link, share, and analyze increasingly voluminous, exquisitely diverse data, or in Rutherford's words, 'stamp collections'. Recall that it was the invention of calculus by Newton and Leibnitz that for the first time allowed precise descriptions of rate of change, and hence fundamental constructs of classical physics such as velocity and acceleration, and the Newton's laws that specified how they related to each other, that helped transform the study of the physical universe from "stamp collecting" to "physics," from a descriptive science into a predictive science. While whether there exist analogs of the simple laws of classical physics for complex biological, cognitive, economic, and social systems might be debatable, that the invention of calculus by Newton and Leibnitz is what made possible the emergence of physics is not.

Mathematics is generally regarded as the language of science. *Algorithms*—precise recipes that describe the relationships between and the processes that operate on the entities that make up the world around us—offer a means for expressing *constructive* mathematics[4]. Algorithms allow us to at least approximate *anything* that is describable, including highly non-linear phenomena that cannot be described using equations that have closed-form solutions. There is a growing recognition that processes of interest in biological, social, and cognitive sciences can be viewed as essentially information processes. Arguably, "applied computer science is now playing the role which mathematics did from the seventeenth through the twentieth centuries: providing an orderly, formal framework and exploratory apparatus for other sciences" (Djorgovski, 2005).

---

[4] Constructive mathematics is distinguished from its traditional counterpart, classical mathematics, by the strict interpretation of the phrase "there exists" as "we can construct" (Bridges and Palmgren, 2013).

This allows us to examine biological, cognitive, and social processes through a *computational lens*, that is, in terms of information processing abstractions (Karp, 2011). Hence, we understand a phenomenon when we have an algorithm that describes it at the desired level of abstraction. Thus, we will have a theory of protein folding when we can specify an algorithm that takes as input, a linear sequence of amino acids that make up the protein (and the relevant features of the cellular environment in which folding is to occur), and produces as output, a description of the 3-dimensional structure of the protein (or more precisely, a set of stable configurations). Examination of natural processes through the computational lens sheds new light on old scientific problems in the respective scientific disciplines. For example, Holland's and Valiant's examinations of biological evolution through the computational lens provide new insights into evolution of complex organisms (Holland, 1975; Valiant, 2009). Roughgarden's work shows how computational complexity sheds new light on the "bounded rationality" of decision-makers (Roughgarden, 2010). Kleinberg's work has provided a new perspective on fundamental questions, e.g., the small world phenomenon (Kleinberg, 2000), in the social sciences.

Examination of a natural process through the computational lens necessarily requires algorithmic or information processing abstractions of the relevant natural entities, relations, and processes. Once such abstractions are created, they become first class computational artifacts in their own right that can be analyzed, shared, and integrated with other related artifacts, contributing to the acceleration of science.

**Computing as a formal framework for science**

The scientific enterprise (See Figure 1), entails acquiring, organizing, verifying, validating, integrating, analyzing, reasoning with, and communicating descriptions of scientific artifacts, namely, experiments, data, models, hypotheses, theories, and explanations associated with natural or built systems lie at the heart of the scientific enterprise. Hence, computing, the science of information processing, offers a powerful formal framework and exploratory apparatus for the conduct of science (Djorgovski, 2005). It also offers the theoretical and experimental tools for the study of the feasibility, structure, expression, and, when appropriate, automation of (aspects of) the scientific process, the structure and organization of collaborative teams, modeling the evolution of scientific disciplines, and measuring the impact of scientific discoveries.

Accelerating science through automation of aspects the scientific process has been a topic of considerable interest in computer science (Duda et al., 1979; deJong and Rip, 1987; Langley, 1981; Langley et al., 1987; Lindsay et al., 1980; Dzeroski et al., 2007; Shrager and Langley, 1990; Valdez-Perez, 1999; Bradley et al., 2001; Glymour, 2004) as well as cognitive science (Klahr, 2000). Intelligent software agents are already widely used in many aspects of scientific activity. However, this work falls short of accelerating most aspects of science (Waltz and Buchanan, 2009).

Recent advances in robotics for data acquisition, data bases and knowledge bases that capture the relevant background knowledge in specific disciplines, open access to large bodies of scientific literature, technologies for connecting resources and experts, and for constructing and sharing scientific workflows have led to a renewed interest in the topic. For example, King et al. (2009) have demonstrated a robot scientist capable of generating and testing hypotheses, and choosing the experiment to try next, to understand the functional genomics of yeast (*S. cerevisiae*). Schmidt and Lipson (2009) have demonstrated a system that discovers compact equations describe complex nonlinear dynamical systems, from observations. These demonstrations suggest the possibility of accelerating science by automating some aspects of the scientific process.

**ACCELERATING SCIENCE: A RESEARCH AGENDA IN COMPUTER AND INFORMATION SCIENCES**

Accelerating science to keep pace with the rate of data acquisition and data processing calls for concerted research efforts that encompass both:

- Development, analysis, integration, sharing, and simulation of algorithmic or information processing abstractions of natural processes, coupled with formal methods and tools for their analyses and simulation;
- Innovations in cognitive tools that augment and extend human intellect and partner with humans in all aspects of science.

In what follows, we elaborate on each of these in turn.

**Algorithmic Abstractions for Accelerating Science**

The success of computational lens in shedding new light on long-standing questions in biological, cognitive, and social sciences is contributing to their transformation from descriptive sciences into predictive sciences. However, in most disciplines, this transformation is far from complete. In many areas, such abstractions are scarce. In others, the abstractions and the hypotheses that they offer have remained untested, at least in part, due in part to the limitations of our instruments of observation and experimentation and in part due to the cost and complexity of the scientific enterprise. In order for a broad range of sciences and scientists to benefit from the use of computational lens in their respective disciplines, there is an urgent need for developing, sharing, analyzing, and integrating computational abstractions or representations of the key entities, relationships, and processes of interest in the respective scientific disciplines. For example, progress in life sciences has been accelerated substantially with the emergence of gene ontology (Ashburner et al., 2000). Much work remains to be done in a similar vein in other scientific disciplines. Of particular interest are system-level, mechanistic, computational models of biological, cognitive, and social systems that enable the integration of different processes into coherent and rigorous representations that can be analyzed, simulated, integrated, shared, validated against experimental data, and used to guide experimental investigations. Such abstractions, coupled with formal methods for their analysis, can provide rich defined modeling languages with precise syntax and semantics that can be analyzed systematically and efficiently for certain properties of interest. For example, a question of interest to a cancer biologist, e.g. whether the up-regulation of genes A and B and down-regulation of gene C could possibly take a cell from a healthy state to a cancerous state can be translated into a reachability query against a model of a cell where the state of the cell encodes the expression levels of the genes. While there has been some progress in developing such abstractions for molecular and systems biology (Priami, 2009; Bernot et al., 2004; Danos and Laneve, 2004; Fisher and Henzinger, 2007), much work remains to be done, especially in relation to formalisms that allow specification of models that take into account uncertainty and variability, as well as couplings across multiple levels of abstraction, e.g., molecules, cells, tissues, organs, organisms. Similar advances are needed in other scientific disciplines. Of particular interest are formalisms for bridging models not only across levels of abstraction, but also, disciplinary boundaries, to allow studies of complex interactions, e.g., those that couple food, energy, water, environment, and people.

**Cognitive Tools for Accelerating Science**

In order for science to keep pace with the rate of data acquisition and data processing, there is an urgent need for innovations in cognitive tools (Saloman et al., 1991) for scientists, i.e., computational tools that leverage and extend human intellect (Engelbart, 1962), and partner with humans on a broader range of tasks involved in scientific discovery (formulating a question, designing, prioritizing and executing experiments designed to answer the question, drawing inferences and evaluating the results, and formulating new questions, in a closed-loop fashion). This calls for for deeper understanding formalization, and algorithmic abstractions of, various aspects of the scientific process; development of the computational artifacts (representations, processes, software) that embody such understanding; and the integration of the resulting artifacts into collaborative human-machine systems to advance science (by augmenting, and whenever feasible, replacing individual or collective human efforts). The resulting computer programs would need to close the loop from designing experiments to acquiring and analyzing data to generating and refining hypotheses back to designing new experiments.

Accelerating science calls for programs that can access and ingest information and background knowledge relevant to any scientific question. As search engines and digital libraries return more articles in response to a query than anyone can read, e.g., Google returns about 3.67 million hits for "cancer biology", there is a need for programs that can read, assess the quality and trustworthiness of, and interpret such information. Machine reading (Etzioni et al., 2006) and information extraction (Niu et. al., 2012) are already active areas of research in computer science that have been successfully applied in the life sciences (Hunter and Cohen, 2006; Mallory et al., 2015) and found their way into commercial technology such as IBM Watson (Ferrucci, et al., 2013). Of particular interest are methods for extracting scientific claims from literature and linking them to supporting assumptions, observations or experiments, answering questions, quantifying uncertainty associated with the answers, etc.

Another active area of research is literature-based discovery (Swanson and Smalheiser, 1997; Smalheiser, 2012), which has had some success in finding new relationships between existing knowledge from literature spanning two or more topics (Cameron et al., 2013). Other work on text analytics has led to powerful methods for understanding the evolution of scientific disciplines (Börner et al., 2004; Sinatra et al., 2015), recommending collaborators (Chen et al., 2011), and choosing experiments to accelerate collective discovery (Rzhetsky et al., 2015). However, many challenges remain, e.g., drawing inferences from disparate collections of literature, and increasingly, scientific databases and knowledge bases that contain information of varying degrees of quality and reliability, tracking the evolution of disciplines, identifying major gaps in scientific knowledge, and areas ripe for breakthroughs.

An emerging area of research focuses on data driven approaches to characterizing, and modeling the evolution of scientific disciplines. For example, the results of a recent analysis of the Physics literature (Sinatra et al., 2015) calls into question the conventional narrative of physics as one of paradigm shifts (Kuhn, 1996), divorced from other sciences, and shows that physics has always been in a constant dialog with other disciplines from mathematics to chemistry and even theology, a dialog that is largely driven by the idea that complex phenomena can be understood in terms of a small number of universal laws. Such analyses could allow us to understand the evolution of scientific disciplines, and the impact of a scientific discovery within and beyond the discipline, and identify unexplored areas that are ripe for investigation.

With the exponential growth in scientific literature, often with conflicting scientific arguments, supported by observations of variable quality and analyses made under differing assumptions, there is a dire need for tools for managing conflicting arguments, tracking changes in the validity of the observations and assumptions that they rely on, and support justifiable conclusions. While there is considerable work on computational argumentation systems (Besnard and Hunter, 2008), much work is needed to develop argumentation formalisms and tools that can help accelerate science. Of particular interest are expressive yet computationally tractable languages for representing and reasoning with scientific arguments, and their uncertainty and provenance.

A shift in emphasis from accelerating data collection and data processing to accelerating the entire scientific process calls for representation and modeling languages with precise formal semantics for describing, sharing, and communicating scientific observations (including measurement models) experiments, data, models, theories, conjectures, and hypotheses. The increasing reliance on cognitive tools requires that the all of these be specified in a form that can be processed by computers; and queries against them be translated into precise computational problems.

Even the relatively mundane task of data collection presents many questions including deciding which variables to measure, why, and how i.e., the instrument to use (if one exists) or to design (if need be). Scientific workflows (Gil et al., 2007; Davidson and Freire, 2008) already provide useful ways to describe, manage, share, track data provenance within, and reproduce complex scientific analyses. Scientific systems are already being used for data analyses in the life sciences (Hull et al., 2006). However, there is a need for languages and tools for describing the measurement process, the data models for describing observations using standard ontologies (when they exist), establishing semantics preserving mappings across data models. There is an urgent need for precise languages and tools for describing experiments, methods for quantifying the marginal utility of experiments, determining the scientific as well as economic feasibility of experiments, comparing alternative experiments, and choosing optimal experiments (in a given context). The same holds for hypotheses, conjectures, theories, scientific workflows, and other scientific artifacts.

Machine learning currently offers one of the most cost-effective approaches to constructing predictive models from data (Ghahramani, 2015; Jordan and Mitchell, 2015) across a number of disciplines including biological sciences (Baldi and Brunak, 2001), brain sciences (Pereira et al., 2009), learning sciences (Romero and Ventura, 2010), biomedical and health sciences (Jensen et al., 2012), environmental science (Hampton et al., 2013), and climate science (Faghmous et al., 2014). For example, in biological sciences, machine learning algorithms are routinely used to build predictors of gene structure (McAuliffe et al., 2004), molecular interactions and interfaces (Xue et al., 2015), and to uncover regulatory interactions between genes (Segal et al., 2003). However, such models are often complex hard for scientists to comprehend, and therefore to use to gain mechanistic insights into the underlying phenomena. Consider for example, a support vector machine using a non-linear kernel that predicts whether a target gene of interest is turned on or off based on the previous states of a few hundred other genes. Such a model, its high predictive accuracy, is virtually useless with regard to helping to uncover the underlying genetic regulatory network. There has been some progress in extracting comprehensible knowledge from complex predictive models (Pazzani et al., 1997). A related topic in which there has been considerable interest has to do with methods for incorporating prior knowledge into machine learning (Heckerman et al., 1995; Fung et al., 2002; Cohen, 2014, Faghmous et al., 2014) as well as cognitive modeling (Tenenbaum et al., 2006). However, there remains a significant language gap between model builders and model users. This language gap presents challenges in exploiting prior knowledge to guide model construction, and in interpreting predictive models produced by machine learning in advancing scientific understanding of the underlying domain. For example, in life sciences, directed labeled graph representations of gene regulatory networks (Honavar, 2013) wherein nodes denote genes and directed edges denote regulatory influences, and the + or - labels on the edges denote the excitatory or inhibitory influences are likely to be much more useful in refining our understanding the underlying process, and in suggesting further experiments, than a black box support vector machine with a complex nonlinear kernel that provides the same prediction. Hence, there is an urgent need for a new generation of machine learning algorithms that that can incorporate prior knowledge and constraints from a variety of sources, e.g., from physics, and produce models are expressed in forms that are easy to communicate to disciplinary scientists.

There has been much progress on methods and tools for integrating data from disparate data sources (Lenzerini, 2002; Doan et al., 2012; Haas, 2015); describing data semantics using expressive yet tractable fragments of logic (Berners-Lee et al., 2001; Baader and Nutt, 2003; Calvanese et al., 2007; Horrocks et al., 1999), and more recently, on the more complex problem of sharing knowledge across disparate knowledge bases (Bao et al., 2009; Cuenca Grau et al., 2008; Kutz et al., 2004; Borgida and Serafini, 2003). Yet many challenges remain, especially as they relate to integration of data and knowledge at different levels of abstraction, differing levels of uncertainty, trustworthiness, etc.

Answering complex questions increasingly requires synthesizing the findings from data from disparate observational and experimental studies to draw valid conclusions. Conclusions that are obtained in a laboratory setting may not hold exactly a setting that differs in many aspects from that of the laboratory. Often, individual studies, for practical reasons e.g., cost, complexity of the studies, focus on the relationship between a selected set of experimental variables and a specific outcome variable. This means arriving at meaningful answers to questions of interest invariably requires synthesize the findings from multiple such studies, carried out under related, but different experimental settings, under possibly different experimental constraints (e.g., experiments that can be performed on a mouse cannot be carried out on human subjects). While causal discovery from disparate observations and experiments is an active topic of research (e.g., Bareinboim et al., 2013), a great deal of work is needed to characterize the precise conditions under which findings of disparate observational and experimental studies can be synthesized, and to develop cognitive tools for synthesizing such findings when it is appropriate to do so.

While we have effective tools to assist scientists in routine aspects of data management and analytics, barring a few proof-of-concept demonstrations (e.g., King et al., 2009), most of the other steps in the scientific process currently constitute rate limiting steps in scientific progress. These include: Characterizing the current state of knowledge in a discipline and identifying the gaps in the current state of knowledge; Generating and prioritizing questions that are ripe for investigation based on the current scientific priorities and the current state of knowledge; Designing, prioritizing, planning, and executing experiments; Analyzing and interpreting results; Generating and verifying hypotheses; Drawing and justifying conclusions; Validating scientific claims; Replicating studies; Documenting studies; Recording scientific workflows and tracking provenance of data and results; Reviewing and Communicating results; Integrating results into the larger body of knowledge within or across disciplines. Hence, accelerating science requires a rich model of the entire scientific process (See Figure 1) as well as deep knowledge of the scientific area under investigation (Honavar, 2014).

Because science is increasingly a collaborative endeavor, we need: sharable and communicable representations and processes, as well as organizational and social structures and processes, that facilitate collaborative science, including mechanisms for sharing data, experimental protocols, analysis tools, data and knowledge representations, abstractions, and visualizations, tasks, mental models, scientific workflows, mechanisms for decomposing tasks, assigning tasks, integrating results, incentivizing participants, and engaging large numbers of participants with varying levels of expertise and ability in the scientific process through citizen science (Gill and Hirsh, 2012; Bonney et al., 2014).

## SUMMARY AND RECOMMENDATIONS

The recent advances in sensing, measurement, storage and communication technologies and the resulting emergence of "big data" offer unprecedented opportunities for not only accelerating scientific advances, but also enabling new modes of discovery. Scientific progress in many disciplines is increasingly driven by advances in our ability to:

- Examine natural phenomena through the computational lens, i.e., using algorithmic or information processing abstractions of the underlying processes;
- Acquire, share, integrate, analyze, and build predictive and causal models from disparate types of data.

However, there is a huge gap between our ability to acquire, store, and process data and our ability to make effective use of the data to advance science. Despite successful automation of routine aspects of data management and analytics, most elements of the scientific process currently constitute rate-limiting steps in the scientific process.

Accelerating science to keep pace with the rate of data acquisition and data processing calls for focused investments in a research program that encompasses both:

- Development, analysis, integration, sharing, and simulation of algorithmic or information processing abstractions of natural processes, coupled with formal methods and tools for their analyses and simulation;
- Innovations in cognitive tools that augment and extend human intellect and partner with humans in all aspects of science. This requires:
  - The formalization, development, analysis, of algorithmic or information processing abstractions of various aspects of the scientific process;
  - The development of computational artifacts (representations, processes, software) that embody such understanding; and
  - The integration of the resulting cognitive tools into collaborative human-machine systems and infrastructure to advance science.

Of particular urgency are investments in:

- Algorithmic abstractions of:
  - The natural entities, relations, and processes of interest in specific scientific disciplines;
  - Formal methods and tools for their analyses and simulation;
  - Formalisms for specification of models that take into account uncertainty, and variability;
  - Couplings across multiple levels of abstraction and spatial and temporal granularity;
- Cognitive tools for:
  - Mapping the current state of knowledge in a discipline and identifying the major gaps;
  - Generating and prioritizing questions that are ripe for investigation based on the current scientific priorities and the gaps in the current state of knowledge;
  - Machine reading, including methods for extracting and organizing descriptions of experimental protocols, scientific claims, supporting assumptions, and validating scientific claims from scientific literature, and increasingly scientific databases and knowledge bases;
  - Literature-based discovery, including methods for drawing inferences and generating hypotheses from existing knowledge in the literature (augmented with discipline-specific databases and knowledge bases of varying quality when appropriate), and ranking the resulting hypotheses;

- Expressing, reasoning with, updating scientific arguments (along with supporting assumptions, facts, observations), including languages and inference techniques for managing multiple, often conflicting arguments, assessing the plausibility of arguments, their uncertainty and provenance;
- Observing and experimenting, including languages and formalisms for describing and harmonizing the measurement process and data models, capturing and managing data provenance, describing, quantifying the utility, cost, and feasibility of experiments, comparing alternative experiments, and choosing optimal experiments (in a given context);
- Navigating the spaces of hypotheses, conjectures, theories, and the supporting observations and experiments;
- Analyzing and interpreting the results of observations and experiments, including machine learning methods that: explicitly model the measurement process, including its bias, noise, resolution; incorporate constraints e.g., those derived from physics, into data-driven inference; close the gap between model builders and model users by producing models that are expressible in representations familiar to the disciplinary scientists;
- Synthesizing, in a principled manner, the findings in a target setting from disparate experimental and observational studies (e.g., implications to human health of experiments with mouse models);
- Documenting, sharing, reviewing, replicating, and communicating entire scientific studies in the form of reproducible and extensible scientific workflows;
- Communicating results of scientific studies and integrating the results into the larger body of knowledge within or across disciplines;
- Collabortating, communicating, and forming teams with other scientists with complementary knowledge, skills, expertise, and perspectives on problems of common interest (including problems that span disciplinary boundaries or levels of abstraction);
- Organizing and participating in citizen science projects, including tools for decomposing tasks, assigning tasks, integrating results, incentivizing participants, and engaging large numbers of participants with varying levels of expertise and ability in the scientific process;
- Cognitive tools for tracking scientific progress, the evolution of scientific disciplines and scientific impact.
- Multi-disciplinary, interdisciplinary, and trans-disciplinary teams that bring together:
  - Experimental scientists in a discipline, e.g., the biomedical sciences, with information and computer scientists, mathematicians, etc., to develop algorithmic or information processing abstractions to support theoretical and experimental investigations;
  - Organizational and social scientists and cognitive scientists to study such teams, learn how best to organize and incentivize such teams and develop a science of team science;
  - Experimental scientists in one or more disciplines, computer and information scientists and engineers, organizational and social scientists, cognitive scientists, and philosophers of science to design, implement, and study end-to-end systems that flexibly integrate the relevant cognitive tools into complex scientific workflows to solve broad classes of problems in specific domains, e.g., understanding complex interactions between food, energy, water, environment, and populations.
- Interdisciplinary graduate and undergraduate curricula and research based training programs to prepare:

- A diverse cadre of computer and information scientists and engineers with adequate knowledge of one or more scientific disciplines to design, construct, analyze and apply algorithmic abstractions, cognitive tools, and end-to-end scientific workflows in those disciplines;
- A new generation of natural, social, and cognitive science researchers and practitioners fluent in the use of algorithmic abstractions and cognitive tools to dramatically accelerate and explore new modes of discovery within and across disciplines.

A research agenda focused on accelerating science can be expected to yield:

- Fundamental advances in multiple areas of computer and information sciences, including, theory of computation, complexity theory, algorithms, formal methods, knowledge representation and inference, information integration, machine reading, software engineering, machine learning, causal inference, multi-objective optimization, argumentation systems, planning, decision making, computational organization theory, robotics, human-computer-robot interaction, among others;
- Cognitive tools that could dramatically accelerate scientific progress, by leveraging and extending the reach of human intellect, and partnering with scientists, including citizen scientists, with a broad range of skills and expertise.

This white paper has sought to articulate a research agenda for developing cognitive tools that can augment human intellect and partner with humans on the scientific process. The resulting new cognitive tools can help realize the transformative potential of big data in many sciences, by dramatically accelerating science. The benefits of accelerating science extend well beyond the scientific community to all of humanity: Precision health regimens that take into account not only one's genetic makeup, but also environment, and lifestyle; Personalized education that optimizes curriculum, pedagogy, etc. to optimize the learning outcomes for each individual; Precision agriculture that optimizes everything from the choice of crops to water and fertilizer use to optimize yield and impact on the environment.

## Acknowledgements

This material is based upon work supported by the National Science Foundation under Grant No. (1136993). Any opinions, findings, and conclusions or recommendations expressed in this material are those of the author(s) and do not necessarily reflect the views of the National Science Foundation. The article has benefited from discussions with and feedback from the broader scientific community, and current and former members of the Computing Community Consortium Council, especially Greg Hager, Cynthia Dwork, Liz Bradley, Klara Nahrstedt, Elizabeth Mynatt, Ben Zorn, Susan Davidson, Susan Graham.

*For citation use*: Honavar V., Hill M., & Yelick K. (2016). *Accelerating Science: A Computing Research Agenda*: A white paper prepared for the Computing Community Consortium committee of the Computing Research Association. http://cra.org/ccc/resources/ccc-led-whitepapers/

This material is based upon work supported by the National Science Foundation under Grant No. (1136993). Any opinions, findings, and conclusions or recommendations expressed in this material are those of the author(s) and do not necessarily reflect the views of the National Science Foundation.